\documentclass[12pt]{article}
\usepackage{graphicx}
\begin{document}
\title{Applications of Small-World Networks to some Socio-economic Systems}
\author{A. S. Elgazzar\\
\textit{Mathematics Department, Faculty of Education}\\
\textit{45111 El-Arish, Egypt}\\
\textit{E-mail: elgazzar@mans.edu.eg}}
\date{}

\maketitle

\begin{abstract}
Small-world networks (SWN) are found to be closer to the real
social systems than both regular and random lattices. Then, a
model for the evolution of economic systems is generalized to SWN.
The Sznajd model for the two-state opinion formation problem is
applied to SWN. Then a simple definition
of leaders is included. These models explain some socio-economic aspects.
\end{abstract}

\section{Introduction}
The diffusion of a new concept (a technology, an opinion, an
information, a disease, ...) through social and economic systems
is a complex process that typically forms waves or avalanches.
Also, it usually displays rich dynamics that is attracting the
interest of many mathematicians and theoretical physicists [1].

A social network has two main properties: clustering and
small-world effect. Clustering means every one has a group of
collaborators, some of them will often be a collaborator by
another person. Small-world effect means the average shortest
person to person (vertex to vertex) distance is very short
compared with the whole size of the system (number of vertices).

Regular lattices display the clustering property only. On the
other hand, random lattices display the small-world effect without
clustering [2]. The concept of SWN introduced in [2,3] has shown
to combine both features. A SWN is a connected ring with some
shortcuts joining between some randomly chosen vertices are added
with small probability $\phi$. Also, this structure combines
between both local and nonlocal interactions. This combination is
observed in many real systems. Then it is a useful concept in
modeling the real systems. We used this concept in modeling
different systems [4-8]. Here our interest is restricted to apply
the concept of SWN to a model for the evolution in economic
systems [8] and to a sociophysics model [7].

\section{A model for the evolution of economic systems in SWN}
A real economic system is a population of agents each of them has
a technological level ($a_i$). Every agent is assumed to interact
with a group of collaborators (neighbours) obtaining payoffs
(profits). The base payoff is assumed to be higher for the higher
technological levels, and the payoffs have to be bounded. Also, if
an agent is too advance or too backwards relative to his/her
neighbours, he/she always has incompatibility costs.

Every agent has to update his/her technological level to obtain
the best payoff. Then the population reaches a state at which each
agent is satisfied with his/her payoff. Then the technological
level of a randomly chosen agent is raised by a quantity $ \Delta
\in (0,1)$. Usually the new technology has a cost. The neighbours
of this agent updates their technological levels according to the
new conditions. They have the possibility to accept or refuse the
new technology. The new technology may diffuse through the whole
population making a uniform front for cheap updating cost. On the
other hand, if the cost is very high, only few agents can modify
their levels. For intermediate cost values, the updating process
continues through the population making an avalanche until another
stable state is reached, and so on. The avalanche's size, $s$ is
the number of agents that updated their levels in a step. It has
been shown that [9] the avalanches size and several aspects of
social and economic systems can be described in terms of power law
distribution as,
\begin{equation}
P (s) \propto s^{-\gamma},\;\;\; \gamma \ge 1.
\end{equation}

There are two updating optimalities: Nash and Pareto defined as
follows:\\
\\
\textbf{Definition 1. }(Nash optimality): An agent selects the
technological level that maximize his own payoff without regards
to his/her collaborators.\\
\\
\textbf{Definition 2. }(Pareto optimality): An agent selects the
technological level that maximize the average payoff of his/her
group (he/she and the collaborators).\\

Towards a theoretical approach to this phenomena, Arenas et al.
[10] introduced the following payoff function:
\begin{equation}
\pi (a_i,a_j)=\left \{\begin {array}{ll}
a_i-k_1(1-{\rm exp} -(a_i-a_j)) & {\rm if} \; a_i\ge a_j,\\
a_i-k_2(1-{\rm exp} -(a_j-a_i)) & {\rm if} \; a_i<a_j,
\end {array} \right .\;\;\;j=i\pm 1,
\end{equation}
where $k_1$ and $k_2$ represent the incompatibility costs
resulting from being too advance or too backwards, respectively.
The payoff of the agent $i, \pi_i$ is
$\pi_i=\pi(a_i,a_{i-1})+\pi(a_i,a_{i+1})$. Depending on this
payoff function, Arenas et al. [10] construct a 1-dimensional
(1-d) model using Nash optimality. But Pareto optimality usually
implies less erratic behaviour than Nash optimality [5]. Also, it
has been proven that the payoff from Pareto updating rule is
generally higher or equal to that from Nash updating rule [11].

Then we have reconstructed the 1-d model of Arenas et al. using
Pareto optimality [8]. It is shown that [8], the dynamics this
model depends only on the quantity $k=k_1-k_2$, which behaves as
the updating cost. For $k\le 2$, the avalanches are of the size of
the whole population corresponding to the case of uniform front.
In the limit $k \to \infty$, the agents behave independently, so
avalanches become of size one. For intermediate values of the
parameter $k$, the system is shown to be critical, see Fig. (1),
where $k=4$. The three regimes are clearly observed in real
systems.

The total payoff of each agent is calculated and compared in two
cases, Pareto optimality and Nash optimality. Applying the same
conditions in the two cases, it is found that the total payoff
from the Pareto optimality is significantly higher than that from
the Nash optimality for all agents. This important result means
that if every one in a population tries to maximize the payoff
(profit, fitness, ...) of his/her group, the personal payoff of
each of them may be more maximized than that of the case when
every one tries to increase his/her own payoff individually.

Then the model is generalized to SWN with $\phi=0.05$. The
shortcuts are fixed beforehand. The updating rule is Pareto. The
vertices that do not have shortcuts behave similar to the 1-d
model. If a vertex $i$ has a shortcutting neighbour ${\rm sc}(i)$,
then its payoff is calculated as follows:
\begin{equation}
\pi_i=\pi(a_i,a_{i-1})+\pi(a_i,a_{i+1})+\pi(a_i,a_{{\rm sc}(i)}).
\end{equation}
In this case the average payoff of its group becomes:
\begin{equation}
\pi_{\rm av}=(\pi_{i-1}+\pi_i+\pi_{i+1}+\pi_{{\rm sc}(i)})/4.
\end{equation}

The application of the model to SWN does not destroy the
criticality of the model for intermediate values of the parameter
$k$. In the limits $k \le 2$ and $k \to \infty$ the model on SWN
behaves as the same as the 1-d model.

The total payoff of each agent was calculated. The 10 \% of agents
who have shortcuts are found to have significantly higher payoff
than the others. This is because of the presence of the third term
in Eq. (3) for calculating the payoff of those agents. This
behavior increases the deviation in payoffs which exists in the
real economic systems. This contradicts the both cases of the 1-d
model (with both Nash and Pareto updating rules), where the
payoffs of all agents are close to each other. Also, it implies
that the increase of collaborators may increase the payoff, if
agents use the Pareto optimality.

\section{Application of the Sznajd sociophysics model to SWN}
The simple Ising model is one of the fundamental concepts of
Statistical Mechanics. Recently, it has been modified to model the
problem of two-state opinion formation [12-14]. Depending on an
old principle that says "united we stand, divided we fall", the
Sznajd model [12] is constructed. Consider a chain of spins
$S_i,\;i=1,2,3,\dots,L$, that are either up ($+1$) or down ($-1$).
Assume that each pair of adjacent spins can affect the state of
their nearest neighbors using the following updating rule:
\begin{equation}
\begin{array}{c}
{\rm if}\; S_i S_{i+1}=+1,\;\;\; {\rm then}\; S_{i-1}=S_{i+2}=S_i,\\
{\rm if}\; S_i S_{i+1}=-1,\;\;\; {\rm then}\; S_{i-1}=S_{i+1},\;
S_{i+2}=S_{i}.
\end{array}
\end{equation}

Simulating this system for long time, where at each time step the
site $i$ is selected randomly, one obtains finally one of the
following three fixed points: $\uparrow \uparrow \uparrow \dots$,
$\downarrow \downarrow \downarrow \dots$, $\uparrow \downarrow
\uparrow \downarrow \dots$ with probability $0.25$, $0.25$, $0.5$,
respectively. Then the model has two possible final states:
dictatorship or stalemate, so no common decision in a democratic
way can be done. Only after introducing a small finite noise, a
democratic decision can be done.

This model is generalized to 2-d [13] obtaining similar fixed
points, except in some variants. Also, it is used to explain the
distribution of votes among candidates in the Brazilian local
elections [14].

We generalized the Sznajd model to SWN [7]. The updating rule is
generalized to include the shortcutting neighbors, if exist, as
follows:
\begin{equation}
\begin{array}{c}
{\rm if}\; S_i S_{i+1}=+1,\;\;\; {\rm then}\;
S_{i-1}=S_{i+2}=S_{sc(i)}\; {\rm (if \; exists)}\\
=S_{sc(i+1)}\;{\rm (if \; exists)}\;=S_i,\\
{\rm if}\; S_i S_{i+1}=-1,\;\;\; {\rm then}\; S_{i-1}=S_{sc(i)}\;
{\rm (if \;
exists)}\;=S_{i+1},\\
S_{i+2}=S_{sc(i+1)}\;{\rm (if \; exists)}\;=S_{i}
\end{array}
\end{equation}
where $sc(i)$ is the shortcutting neighbor of the $i$-th vertex,
if exists.

Beginning with a totally random initial state, the system always
reaches one of two fixed points all up or all down with equal
probabilities. Also, the effects of the initial concentration of
the up and down spins is studied [7]. Although the stalemate fixed
point disappeared in this model, the model always evolves to
limiting dictatorship fixed points. So no democratic decision can
be taken in this system. This point is improved by introducing the
concept of a leader.

A leader is usually a strong person who completely trusts in
his/her opinion. Also, he/she is always capable in influencing
somebody to follow his/her thought without changing. As a simple
definition for a leader, we define a leader as a person who does
not change his/her opinion and has the capability of influencing
one of his/her nearest neighbors to accept the same opinion as the
leader.

We randomly choose two persons to be leaders, one for the up
direction and the other to the down direction. The updating rule
is the same as Eq. (6). The same numerical investigation is
reapplied to this modified model. No fixed point is observed in
this system. The time evolution of the magnetization,
($M=\frac{1}{L} \sum_{i=1}^{L} S_i$) is drawn [7]. Sometimes the
majority of the population follow the up direction, and another
times the majority follow the down direction. There is no
dependence on the initial concentrations for the up and down
followers. There is no periodicity observed. Also, this behavior
is independent of the lattice size.

Because of the disappearing of the dictatorship fixed points, a
decision can be done in a democratic way. This behavior is clearly
observed in many local elections in many democratic countries. For
example, sometimes the majority is for the Democratic party and
others the majority is for the Republic party, without reaching a
state at which $100 \%$ of the population follow a certain
opinion. Also, this behavior is not periodic. Similar behavior is
observed from both 1-d and 2-d versions of Sznajd model after
including some noises.

An interesting question is what will happen if one of the two
leaders is killed or forced to slough his leadership by any way?
In this case the system evolves into a single dictatorship fixed
point following the opinion of the leader that is still alive,
independent of the concentrations of each opinion at the killing
process. This explains why dictators usually kill the active
objectors members, and gives a warning to leaders, they have to
prepare some leaders for the future. The model is investigated
using time series analysis [7].

\section {Conclusions}
In conclusion, SWN is a good description for real social networks.
It allows some simple mathematical models to display some
interesting socio-economic aspects.\\
\\
\textbf{Acknowledgements}\\
\\
I thank E. Ahmed, J. A. Holyst, H. Kantz and D. Stauffer for
helpful discussions and comments.

\newpage

\begin{figure}
\begin{center}
\includegraphics[angle=-90, width=0.5\textwidth]{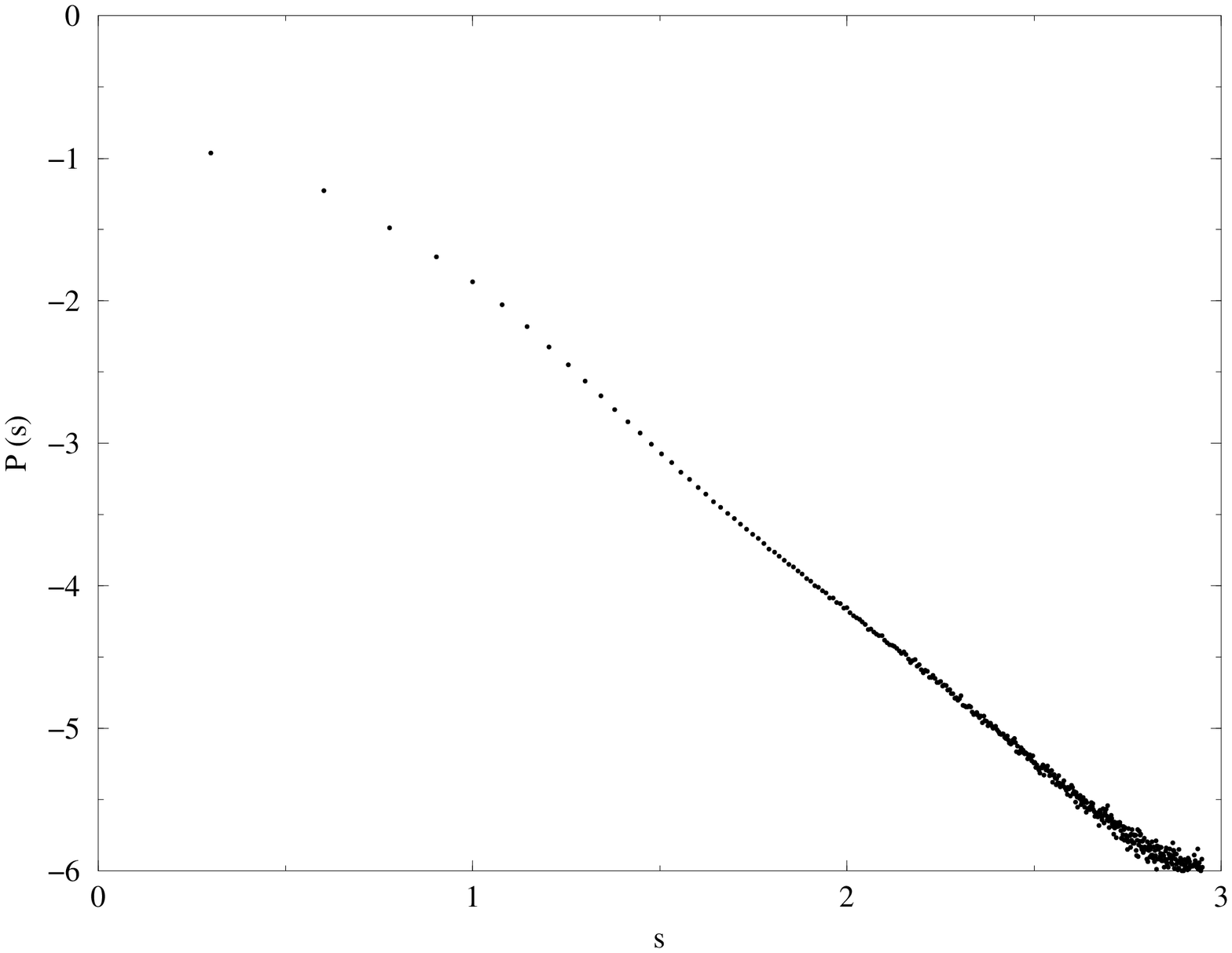}
\caption{A log-log plot for the probability distribution function
of avalanches' sizes $P(s)$ versus the avalanches' sizes $s$ for
the 1-d model using Pareto optimality. A power law behaviour is
obtained with exponent $\gamma= 1.99$, for $k=4$. The results are
averaged over ten independent runs on a 1-d lattice of size $1000$
updated for $6 \times 10^6$ time steps.}
\end{center}
\end{figure}

\end{document}